\documentclass{article}
\usepackage[utf8]{inputenc}
\usepackage{xcolor}
\usepackage{graphicx}
\usepackage{units}
\usepackage{physics}
\usepackage{amssymb,latexsym,amsmath,txfonts,color,graphics}
\usepackage[colorlinks,linkcolor=blue,urlcolor=blue,citecolor=black,
plainpages=false,pdfpagelabels,breaklinks]{hyperref}
\usepackage{soul} 

\usepackage{microtype}
\usepackage{authblk}

\newcommand{\lra}{\leftrightarrow}
\newcommand{\ita}{\textit}
\newcommand{\qst}{$\mathfrak{Q}$}
\newcommand{\mcal}{\mathcal}

\newcommand{\igual}{:=}

\newtheorem{dfn}{Definition}[section]

\title{Quantum identity, content, and context: from classical to non-classical logic}

\author{J. Acacio de Barros\footnote{Corresponding author. Email: barros@sfsu.edu}}
\affil{School of Humanities and Liberal Studies, San Francisco State
University, 1900 Holloway Ave., San Francisco, CA, USA}
\author{Federico Holik}
\affil{Universidad Nacional de La Plata, Instituto de F\'{\i}sica
(IFLP-CCT-CONICET), C.C. 727, 1900 La Plata, Argentina}
\author{D\'ecio Krause}
\affil{Department of Philosophy, Federal University of Santa
Catarina, Florian\'opolis, Brazil, and Graduate Program in Logic and
Metaphysics, Federal University of Rio de Janeiro, RJ, Brasil.}

\date{\today}

\usepackage{graphicx}

\begin{document}

\maketitle
\begin{abstract}
    In this paper, we discuss content and context for quantum properties. We give some examples of why quantum properties are problematic: they depend on the context in a non-trivial way. We then connect this difficulty with properties to the indistinguishability of elementary particles. We argue that one could be in trouble in applying the classical theory of identity to the quantum domain if we take indiscernibility as a core and fundamental concept. Thus, in considering indistinguishability as such a fundamental notion, it implies, if taken earnestly, that one should not apply standard logic to quantum objects. Consequently, we end with a discussion about novel aspects this new mathematics brings and how it relates to some issues associated with the quantum world's ontology and the classical limit. We emphasize that, despite several different ways of questioning classical logic in the quantum domain, our approach is distinct. It involves one of the core concepts of classical logic, namely, identity. So, we are in a different paradigm from standard quantum logics.

\end{abstract}

\section{Introduction}
Quantum mechanics (QM) is a very successful theory. It is also a
strange theory. Though QM can calculate many experiments' outcomes,
there is no consensus about what quantum models tell us about the
microscopic world. In other words, it is unclear what is the
relationship between QM and metaphysics. In this paper, we examine
one particular aspect of the quantum world: quantum particles seem
to lack identity.

Under certain circumstances, two quantum systems of the same kind
(e.g., two electrons) become utterly indistinguishable by any
empirical means. However, the lack of identity comes from more than
just the impossibility of distinguishing between two quantum
particles (e.g., two electrons). It derives from the fact that
nothing changes when we permutate two identical quantum particles,
contrary to what happens in the classical world. This invariance by
permutation is at the core of the Bose-Einstein and Fermi-Dirac
statistics. In this way, the standard interpretation of the theory
assumes indistinguishability. Here, we argue that
indistinguishability is an essential concept in quantum theories
(both non-relativistic and quantum field theories).
Indistinguishability should be thought of as at the same level as
celebrated quantum concepts, such as superposition (in particular,
entanglement), contextuality, and nonlocality.

Some philosophers and physicists are reluctant to admit that
indistinguishability, also known as indiscernibility, plays a
salient role in quantum physics's ontology. Perhaps, this reluctance
comes from the notion that indistinguishability can be simulated
within a ``classical'' mathematical setting, as we shall see below.
However, we find this argument weak for several reasons.

First, just because we can do something does not mean that this is
the best approach. Consider, for example, the geometry of curved
spaces. We may describe a curved space using Riemannian geometry,
where Euclid's postulate of parallel lines is not valid.
Alternatively, we can describe the same space by embedding it in a
higher-dimensional space and keeping Euclid's postulates. Both
approaches yield the same results: all geometry theorems on the
curved space are valid in both descriptions. However, one requires a
more complicated ontological structure with extra dimensions. Should
we make our ontology unnecessarily complicated to accommodate our
prejudices? We believe not.

Second, when someone is interested in a theory's foundations, the
underlying logic and mathematics become fundamental. We should not
do away with an ontological feature because we can use a
mathematical trick to describe it. Instead, we argue that the
mathematical formalism used to cope with quantum systems'
description should consider the ontological features that one aims
to describe. Therefore, as we discuss below in more detail, it is
crucial to develop a mathematical framework that accommodates
indistinguishability in a natural way. In fact, we cannot cope with
a contradictory theory (as some claim is Bohr's theory for the atom,
yet this is disputable -- see the discussion in \cite{vickers})
within a ``classical'' framework such as in the mathematics
developed in a standard set theory such as the ZFC system, which we
presuppose here.\footnote{ZFC is the Zermelo-Fraenkel set theory
with the Axiom of Choice. The reader can think of it as formalizing
the intuitive notion of a set one learned in our math classes.}

Thus, we wish to pursue a metaphysics of non-individuals. In this
metaphysics, quantum entities\footnote{The notion of a
\textit{quantum object}, or quantum \textit{system}, varies from one
approach to another. In orthodox quantum mechanics, we have
particles and waves. In the quantum field theories, the basic
entities are fields, and particles arise as particular
configurations of the fields. Our claims in this paper apply to both
particles and fields.} (here, quantum objects, independently of
their proper characterization) are seen as not following the
standard notion of identity (to be discussed below). Therefore, we
need to change logic and mathematics, unless we accept the
physicists' usual way of impersonating them within classical
frameworks. These entities need to be considered in most cases as
\textit{absolutely} indiscernible, something forbidden in the
classical settings.\footnote{For a defense of the non-individuals
view, see \cite{kraarebue20}.}

Nevertheless, the interpretational problem does not end with the
indiscernibility of quantum objects. Indistinguishability is not the
only mystery of quantum theory. The ontological status of properties
of these objects is also relevant. Quantum properties are tricky,
and if we are not careful about how we deal with them, we may reach
contradictions. These contradictions arise from considering the
possible results of multiple (and incompatible) experiments over the
same system. As we have stated elsewhere \cite{debarros}, we
\textit{never} perform the same experiment twice. What we do is take
a similar experiment, so similar as to be \ita{indistinguishable}.
Since experiments are associated with properties, we should consider
indiscernible properties also. These indistinguishable properties
are also forbidden by classical logic. We need to go outside of
standard mathematics and use a different mathematical (and logical)
setting as, for example, \textit{quasi-set theory}, to be sketched
below. Given that we need to recreate indiscernible properties and
systems, it is natural to use a mathematical setting that
incorporates indistinguishability as a primitive notion right from
the start.

This paper is organized as follows. In Section \ref{sec:context}, we
first discuss the role of context and content in classical and
quantum physics. These two concepts play an essential role in the
difficulties physicists and metaphysicists face concerning quantum
properties. In Section \ref{sec:identity}, we consider the concepts
of identity and indiscernibility and how they are connected.
Identity is a difficult concept, and we explore it both as it is
connected to classical physics and indiscernibility in logic. This
discussion opens up to our investigations outlined in Section
\ref{seq:connecting-identity}. In this section, we argue that by
intimately connecting identity to context, we can solve some
puzzling aspects of quantum physics. Finally, in Section
\ref{seq:logical-approaches}, we outline how to change mathematics
to allow for the existence of indiscernibility as a fundamental and
primitive concept. This mathematics, grounded on quasi-set theory,
captures the idea that quantum objects are indistinguishable and
lack a classical identity. As a bonus, we included in Section
\ref{seq:formulating} somewhat more detailed mathematical
explanation of the structures discussed in Section
\ref{seq:logical-approaches}. We hope the interested reader will
find this useful, but this section can be skipped by those readers
not seeking further mathematical details. We end the paper with some
final remarks, conclusions, and perspectives.

This article is written for a layperson with a strong mathematical
background. The reader is assumed to know enough mathematics to be
comfortable with logic, set theory, and orthodox quantum mechanics.
It should be remarked that a paper dedicated to foundations and
aimed at a general reader requires many caveats, since the delicate
aspects can be quickly passed unsuspected. We try to warn the reader
about those details in between the text or in the footnotes. We ask
the reader's forgiveness in advance for the numerous footnotes.


\section{Content and context in quantum and classical physics}\label{sec:context}

The idea of content and context comes from linguistics, specifically
semantics and pragmatics. Nevertheless, physics has
straightforwardly borrowed those concepts. This section will discuss
how content and context translated from linguistics to physics,
focusing on quantum mechanics. We organize this section in the
following way. First, we concisely review the concepts from
linguistics. Then we explore how content and context show up in
classical semantics theories. Our discussion should not be thought
of as a detailed scholarly review of the linguistic literature on
content and context, as this topic is the object of intense research
in philosophy of language and linguistics for more than a century.
Instead, we present a subset of linguistics that is relevant to
physics. With that in mind, we follow our linguistics discussion by
examining some physics examples. We see that contents may present
context-dependency in both classical and quantum physics. However,
we also argue that the context-dependency in quantum physics is
different.

Let us start with the concept of content. Roughly speaking, semantic
content refers to the meaning of a sentence.\footnote{We shall
assume this without further discussion, but things are not as
straightforward as it may appear. Meaning means ``meaning for
someone,'' and there is no meaning \textit{tout court}. Yuri Manin,
in his great book \cite[pp. 34ff]{manin77} mentions the case of Lev
Alexandrovich Zasetsky, who suffered a brain injury in battle.
Zasetsky could write sentences with meaning, such as ``An elephant
is bigger than an ant," and know that it is true (semantically well
defined). But his illness impeded him to understand the meaning of
the terms  ``ant'' and ``elephant.'' He had semantics and truth, but
not meaning.} Consider the following statement, made by Vera's
friend, Alice:
\begin{enumerate}
\item[L1.] Vera had a bad date.
\end{enumerate}
Sentence L1 can be seen as a proposition referencing to an object.
Assuming the correspondence theory of truth,\footnote{We also
sustain that the correspondence theory of truth,  for instance that
treated by Tarski, is not suitable for the empirical sciences, but
this is something to be developed in another opportunity; here we
take the standard view.} its truth value  requires some metric,
likely subjective, of what constitutes a ``bad date.'' However, once
such a metric exists, one could infer L1's truth value. The
truth-value of L1, therefore, lies on its semantic content. In other
words, a sentence's semantic content can be thought of as a function
that takes the sentence and outputs a truth-value.

Context, on the other hand, is the idea that some statements and
utterances depend on the circumstances surrounding it, such as time,
place, speaker, hearer, and topic, to name a few. For example,
Alice's claim that ``Vera had a bad date'' has different meanings
depending on whether their conversation revolved around the fruits
of the \textit{Phoenix dactylifera} or romantic engagements. The
context alters the meaning and the functions that take the content
to truth values.

However, context does not alter meaning only. Consider the case of
indexicals. The statement ``Acacio is hungry now'' is contingent on
when it is uttered and on the particular subjective satiety state of
the person named ``Acacio.'' In a sense, its meaning does not
change. Its referent, Acacio, is the same (assuming we are talking
about the same person, one of the co-authors of this paper), the
concept of hunger is invariant, and the meaning of now as the
present moment is maintained. However, its truth value is variable.
As we write this paragraph, it is false, as Acacio just had lunch.
However, the same statement was right about an hour ago. It will be
true again several hours from now, even though its meaning is
seemingly unchanged.

To summarize, sentences have meanings given by their semantic
contents. Sometimes the meanings are context-dependent, as in the
case of dates. However, other times, their truth-values vary with
context, whereas their meanings seem to do not. We shall see that
physics has some correlates to those ideas.

Let us start with classical physics. A physically-relevant
proposition about an object is something empirically measurable. For
example, we can have the following statement:
\begin{enumerate}
\item[P1.] A billiard ball's kinetic energy is between $0.1~ kg\cdot m^2/s^2$ and $0.2~kg\cdot m^2/s^2$.
\end{enumerate}
Similarly to linguistics, P1 has a meaning: if we measure the
kinetic energy of a billiard ball, perhaps by measuring its mass and
speed and inferring the energy, we find it to be in a certain range.
Its meaning is given by an accompanying experimental procedure that
yields a truth-value to the sentence. As importantly, this
truth-value also corresponds to the idea that the billiard ball, if
P1 is true, has a specific property: its kinetic energy.

As in linguistics, P1 refers to a subject (the billiard ball) and a
truth-value associated with some meaning-constructing procedure (the
experiment). Accordingly, we can think of any physics experiment as
observing a physical system's property. This property itself has an
associated proposition whose truth-value is assessed by an
experiment. So, in a certain sense, properties of physical systems,
such as temperature, momentum, energy, present an analogy with
contents.

We may take the meaning of a statement as which experiment can yield
a truth-value to it. Consequently, expressions such as P1 attach a
property to a physical object. Of course, the property is the
statement itself, and the experiment is a way to determine its
truth-value.  To summarize, the properties of a physical system are
the content of the propositions.

What about context? Are classical properties context-dependent? Let
us examine an example from 18th-century physics. A group of Italian
researchers in the 1700s, known as the Experimenters, did not
differentiate between heat and temperature but combined both
concepts into one (Wiser and Carey, 1983). This combined concept of
heat and temperature led to some puzzling results. For instance, the
Experimenters wondered about examples such as the following. Imagine
we heat a 2 kg piece of iron and immerse it in a container with room
temperature water, subsequently measuring the water's temperature.
Now, imagine that instead of iron, we use 2 kg of a 3:1 mixture of
nitric acid (1.5 kg) and tin (0.5 kg), immersing it in water, as we
did with the piece of iron. It was surprising to the Experimenters
that even when the mixture of tin and acid was not as hot as the
iron, the latter would not raise the water's temperature as much. If
both objects, iron and mixture, had the same amount of ``hotness,''
why would they increase the water by different levels of
``hotness?''

The answer to the above puzzle is straightforward in contemporary
physics, as we distinguish heat and temperature. Because of this
distinction, we can measure how much heat a substance holds as their
temperature increases: what physicists call specific heat. With this
concept, we can measure that iron has a specific heat of 0.44 J/kg
K. In contrast, the specific heat of a 3:1 mixture of nitric acid
and tin is 1.34 J/kg K. This means that for every one-degree
increase in temperature, the amount of heat held by the 2-kg block
of iron increases by 0.88 Joules and by 2.64 Joules for the 2-kg
tin-nitric acid mixture. In other words, at the same temperature,
the mixture holds three times the amount of heat as the iron.
Because the Experimenters had a single concept of heat and
temperature, they could not even investigate the concept of specific
heat, nor could they understand the puzzle.

Let us examine the example above from a slightly different
perspective. Imagine we are observing a student who does not
distinguish temperature from heat (as the Experimenters) and thinks
of both as the smorgasbord concept ``hotness.''  Consider the
following propositions observed to be empirically true for a
specific experimental setup involving three objects: $X$, $Y$, and
$W$ (as for instance $X$ is iron, $Y$ is the mixture of nitric acid
and tin, and $W$ is water as in the example above).

\begin{description}
\item[$A$:] If X has more heat than Y, then W will have a high temperature.
\item[$B$:] If X has a higher temperature than Y, then W will not have a high temperature.
\end{description}
Both propositions $A$ and $B$ can be true if we carefully chose $X$
and $Y$'s masses, heat capacities, and how we define statements such
as ``low temperature,'' ``high temperature,'' and so on. However,
let us rephrase $A$ and $B$ in terms of the student's hotness
concept. We now have two new propositions, $A'$ and $B'$:
\begin{description}
\item[$A'$:]
If $X$ has more hotness than $Y$, then $W$ will have high hotness.
\item[$B'$:] If $X$ has more hotness than $Y$, then $W$ will not have high hotness.
\end{description}
$A'$ and $B'$ cannot be both true, as they are contradictory. The
contradiction comes here from identifying heat and temperature as a
single concept: hotness.

There is an obvious, albeit silly, solution to this contradiction.
The student might say, \textit{ad hoc}, that ``hotness'' in the
context of an experiment observing $A'$ is different from experiment
$B'$, so they are not the same statement. To save their hotness
concept, the student makes things unnecessarily more complicated
than they need to be. As more experiments pile up, the more contexts
and the more complicated their theory becomes. Furthermore, such a
move would lead to a theory incapable of making good predictions in
different situations.

Of course, this is not what scientists usually do. Scientists try to
find appropriate ways to describe a physical system that does not
lead to contradictions or context dependency. In the hotness case,
they realized that differentiating between heat and temperature was
consistent and allowed for predictions and explanations of thermal
phenomena. When faced with contradictions, scientists realized that
the best approach is to face them and figure out ways to rethink our
theories or experiments without resorting to context-dependency.

The above example is interesting for historical reasons, but it also
illustrates a type of explicit contextuality. In the physics
literature, this explicit contextuality is called direct influences
\cite{ehti2017} or signaling \cite{popescu_quantum_1994}. When the
student ``explained'' the differences between $A'$ and $B'$ as
context-dependent, he thought of explicit contextuality. Explicit
contextuality manifests when there is a direct contradiction between
two statements or results, such as the contradiction between $A'$
and $B'$. When this happens, scientists recognize a problem and try
to solve it, as with the development of the concepts of heat and
temperature.

Let us now move from classical to quantum physics. Quantum physics,
as far as we know, forbids any type of properties that exhibit
direct influences, i.e., signaling.  However, it allow another type
of context-dependency (or contextuality): implicit contextuality. In
the technical literature, this is called simply ``contextuality.''
We call it implicit contextuality to emphasize its contrast with
contextuality due to direct influences. From now on, when we talk
about contextuality, we will refer solely to implicit contextuality.

To understand contextuality in quantum physics, let us consider
another example \cite{specker1975}. Imagine a Simon-like-game device
with three buttons (instead of the usual 4). Each button on this
device, when pushed, randomly emits red or green light. Turns
consists of multiple trials, where after observing their behavior,
the player can try to predict how each button will lit. For each
trial of this game, the player can push at most two buttons at the
same time, for as many times as they want, and in any combination of
the three buttons they wish. If all three buttons are pushed at the
same time, no light is emitted. To win the turn, the player needs to
correctly guess what color the unpressed button would light in their
last trial.

Let us consider a simple non-contextual example for this game.
During her turn, Alice notices the following.
\begin{itemize}
    \item For trials when she only presses one key, they seem to yield either color randomly. In other words, if Alice presses $X$, 50\% of the time he observes green and 50\% red.

    \item For trials when Alice presses $X$ and $Y$, she also gets 50\% for each color for $X$ or $Y$, and the two colors are the same;

    \item For trials when Alice presses $X$ and $Z$, she also gets 50\% for each color for $X$ and $Z$ trials colors are opposite;

    \item For trials when Alice presses $Y$ and $Z$, she also gets 50\% for each color for $Y$ and $Z$ trials colors are also opposite.
\end{itemize}
 So, after realizing that, if Alice presses $X$ and $Y$ and obtain ``red'' for both, she could logically infer that $Z$ would be ``green.'' This is because $Z$ has the opposite color of both $X$ and $Y$. Guessing ``green'' would win Alice the turn.

Now, imagine that in another turn, Bob starts prodding different
combinations of pairs of $X$, $Y$, and $Z$, and observes the
following.
\begin{itemize}
    \item For trials when Bob only presses one key, they seem to yield either color randomly. In other words, if Bob presses $X$, 50\% of the time he observes green and 50\% red.

    \item For trials when Bob presses $X$ and $Y$, he also gets 50\% for each color for $X$ or $Y$, but the two colors are the opposite;

    \item For trials when Bob presses $X$ and $Z$, he also gets 50\% for each color for $X$ and $Z$ trials colors are opposite;

    \item For trials when Bob presses $Y$ and $Z$, he also gets 50\% for each color for $Y$ and $Z$ trials colors are also opposite.
\end{itemize}
In other words, when two buttons are pushed simultaneously, they
randomly emit red or green light, but in opposite colors. This
example exhibits implicit contextuality. To see this contextuality,
imagine we start with $X$ emitting green and $Y$ red. Bob can reason
that if he pushed $X$ and $Z$ instead, then $Z$ would be red.
However, he could also argue that if he pushed $Y$ and $Z$, since
$Y$ was red, $Z$ would be green. Here we reach a logical
contradiction: $Z$ would be both red and green, and impossibility in
the game. To avoid such contradiction, we need to either assume that
$Z$ has no possible color, or that its color changes with the
``context'' of being seen with $X$ or with $Y$. To convince
themselves that $Z$ changes with which other buttons it is pushed,
we urge the readers to think about possible mechanisms that could
yield the outcomes we described. The reader will quickly see that
any mechanism that generates the outcomes for $X$ and $Y$ needs to
be physically different from one generating $X$ and $Z$ (for an
example using a firefly in a box, see
\cite{de_barros_negative_2016}).

The above example of contextuality is contrived. But contextuality
shows up in quantum mechanics. One such example comes from the
Greenberger-Horne-Zeilinger state \cite{ghz}, also known as GHZ.
Without going into the details of where the following relations are
derived, the GHZ state predicts the existence of six observable
properties, $X_1$, $X_2$, $X_3$, $Y_1$, $Y_2$, and $Y_3$, satisfying
the following properties. First, the properties $X_i$ and $Y_i$ take
values $+1$ or $-1$. Second, whenever we observe each of those
properties separately, they look completely random, i.e., their
average value is zero. The same is true for when we observe them in
pairs: they look completely uncorrelated. Third, we can observe them
in triples, and when we do, we see the following relationship
between the triplets.
\begin{align}
 Y_1 Y_2 Y_3 &=1,\label{eq:ghz_y} \\
 Y_1 X_2 X_3 = X_1 Y_2 X_3 = X_1 X_2 Y_3  &=-1.\label{eq:ghz_xy}
\end{align}
The above correlations are experimentally observed
\cite{de_barros_inequalities_2000,bouwmeester_observation_1999}.
Finally, we cannot observe all six properties at the same time. In
fact, we can only observe at most three of them simultaneously. For
example, quantum mechanics forbids us to see $Y_1$, $X_1$,$X_2$, and
$X_3$ at the same time. Contextuality manifests in a similar way as
the previous three-variable example.

To see how contextuality manifests itself, let us assume that the
six properties are \emph{not} contextual. Then, we can use
\eqref{eq:ghz_y} and \eqref{eq:ghz_xy} and write the following.
\begin{equation}
(Y_1 X_2 X_3)(X_1 Y_2 X_3)(X_1 X_2 Y_3)  = (-1)(-1)(-1)  = -1.
\end{equation}
But we can regroup the above product, and get
\begin{equation}
Y_1 Y_2 Y_3(X_2 X_3)(X_1  X_3)(X_1 X_2 )  = Y_1 Y_2 Y_3 (X_2^2)
(X_1^2) (X_3^2).
\end{equation}
However, because $X_i$ is $\pm 1$ valued, their square is $1$, i.e.,
$X_i^2=1$. Therefore, it follows that
\begin{equation}
\label{eq:contradiction}
    (Y_1 X_2 X_3)(X_1 Y_2 X_3)(X_1 X_2 Y_3) = Y_1 Y_2 Y_3.
\end{equation}
But this is a mathematical contradiction! The first term in the
above equation is $-1$ whereas the second term is $+1$, and
\eqref{eq:contradiction} is telling us that $1=-1$.

Where is the contradiction coming from? It does not come from a
mathematical mistake, but from an assumption of non-contextuality.
When we wrote that $X_1^1=1$, we implicitly assumed that $X_1$
observed together with $Y_2$ and $X_3$ is the same as when observed
together with $X_2$ and $Y_3$. This turns out to be false. If we,
instead, call each $X_i$ by a different name depending on the
context, no contradiction is obtained. What happens in quantum
mechanics is similar to the simple color game we discussed before.

The reader may now be thinking about whether we could make a move
similar to the contextual classical case. Namely, can we redefine
properties such that no such kind of contradictions arise in quantum
physics? The answer is yes. Unfortunately, there are many different
ways to do so, and there is no consensus among the physics community
as to which answer is even acceptable. So, let us end this section
with two possible ways around this contradiction.

One move is to assume that properties depend on the context. This is
the idea behind Bohm's interpretation of quantum mechanics
\cite{bohm_suggested_1952,holland_quantum_1995}. In Bohm's theory,
the famous duality wave/particle is resolved by assuming both wave
and particle existence. The wave fills out the whole of space, and
this wave guides the particle. How the wave directs the particle in
one direction or another depends on its form. For example, in the
two-slit experiment, the wave goes through both slits
simultaneously, and due to its interference pattern, it guides the
particle toward certain areas and away from others. The result is
different if one or two slits are open \cite{holland_quantum_1995}.
Since the wave depends on the context dictated by the physical
experiment, Bohm's theory tells us that particles' reality and their
properties are contextual. However, Bohm's theory presents a
problem: for two or more particles, their waves are affected by
their corresponding particle's positions. This theory implies the
existence of instantaneous interactions between physical systems.
Instantaneous interactions present a difficulty to the causal
structure in Bohm's quantum world. As Einstein showed, to have cause
and effect, we cannot have instantaneous interactions. This
difficulty between Bohm's theory and Einstein's special relativity
is the main reason for many physicists to reject it.

Bohm's theory gets into trouble with special relativity because it
assumes that properties exist, whether we choose to measure them or
not. When we measure, we affect the wave function and, consequently,
the physical system. However, the property exists independent of an
observer. In other words, Bohm's theory assumes that reality exists,
whether we observe it or not.

Another possible solution to the problem of contextuality,
particularly to contextuality at a distance (also called
non-locality), is to assume quantum properties do not have values
before a measurement and that the measurement process ``creates''
such values. This position was held by Bohr and is the core of the
Copenhagen interpretation of quantum mechanics
\cite{jaeger_entanglement_2009}. In this interpretation, saying that
an electron has spin $\hbar/2$ in the direction $z$ is meaningless
unless we perform a measurement of spin in the direction $z$ and
find it to be $\hbar/2$. However, before such a measurement, we
cannot say anything about the spin. Furthermore, when we afterward
make a measurement of spin in an orthogonal direction, say $x$,
because $z$ and $x$ spins are incompatible (i.e., cannot be measured
simultaneously), we cannot say anything anymore about the spin in
the $z$ direction; such ``property'' becomes meaningless. So, Bohr
solves the problem of properties in quantum physics by merely
denying their ``existence'' prior to a measurement.

We shall not cover all possible solutions to defining properties in
quantum theory, as they abound. We just wanted to present to the
reader two possible paths on how to deal with it and emphasize that
the choices we have are not necessarily great. In Bohm's theory, we
need to re-think the concepts of causality and space-time, two
well-established tenets of special relativity, to accommodate
faster-than-light signaling. In the Copenhagen interpretation, it
becomes problematic to talk about a reality independent of a
measurement apparatus (and the observer behind it). Either solution
present metaphysical difficulties that have troubled physicists for
more than a century. These puzzles all boil down to the problem of
having properties that depend on the context.

To summarize, in this section, we discussed the idea of content and
context. We started with its origins from linguistics and presented
an interpretation that allows us to apply these concepts to physical
phenomena. We saw that contextual dependencies appear in classical
physics, but they are resolved by resorting to reinterpretations and
refinements of the theory. We then discussed another contextual
dependency that appears in quantum mechanics, such as the GHZ-state
example. We then presented some of the proposed solutions to the
problems and their corresponding metaphysical issues. In the
following sections, we will show that those issues are intimately
related to the concept of identity in the quantum world.

\section{Identity and indiscernibility}\label{sec:identity}

Identity is an old and difficult notion to be dealt with. Usually,
the discussions have focused on personal identity and identity
through time. Here, we shall be concerned with particular
applications of this notion to the identity of objects and
properties. By ``identity of objects,'' or \textit{individuals} as
we prefer to call them,\footnote{The word ``individual,'' according
to the Oxford Online Etymological Dictionary, means ``one and
indivisible.'' Hence our preference for the term. However, as it is
common practice, we relax the idea of `indivisible' and keep
``one,'' adding that it can always be distinguished in other
contexts, at least in principle, from any other individual as being
\textit{that} individual. This distinguishability cannot occur with
quantum entities, even those trapped by some device.} we mean
identity of those entities which are dealt with by the theories of
physics\footnote{The standard quantum formalism is developed within
a mathematical structure called ``Hilbert-space formalism,''
although there are alternatives (\cite{styer} mentions \textit{nine}
different ways of developing orthodox quantum mechanics). }. For a
more detailed discussion about the origins of the term ``object,''
see \cite[pp.13ff]{tor86}; here we review briefly some aspects of
the argumentation given in \cite[Chap.1]{french_identity_2006}.

We have an intuitive idea of what it means to say that two objects,
or individuals, are identical: they are \textit{the same}. However,
to say this is to say nothing, for we also do not know what is to be
``the same,'' something reported equivalent to identity. Thus, we go
to the opposite side: we judge individuals as being
\textit{different} and, therefore, \textit{not identical}, hence
\textit{not the same}. Nevertheless, in virtue of what should
individuals be different? Usually, we look for their differences;
although quite similar, two peas show differences, maybe some small
scratch or a slightly different color. At least, that is what we
tend to think.

Still, in virtue of what two objects would be different? Are they
so? Is it possible to have two (or more) objects perfectly alike,
with no differences at all? Put in other words, what makes an object
an individual, distinct from any other? Is there some Principle of
Individuation we can use to specify an individual's individuality?
Theories of \textit{individuation} are generally divided up in two
main lines: \textit{substratum theories} and \textit{theories of
bundles of properties}. According to the first group, beyond the
properties of an object and the relations it can share with others,
there is \textit{something more}, something Locke described as ``I
don't know what" \cite[Book I, XXIII, 2]{locke}. This notion and the
related ones (such as haecceities and thisness)\footnote{There are
peculiarities in using these terms, but broadly speaking, all refer
to something beyond an individual's properties.} were discarded in
favor of bundle theories of individuation. Bundle theories say that
there is nothing more to an object than the collection of its
properties (encompassing relations). Nevertheless, if in the
substratum theories one could say that what distinguishes an object
from another is its substratum (or something like that), in bundle
theories, many discussions have appeared concerning the possibility
of two objects having the same collection of properties. Can they
have the same collection of properties? If not, why not? Of course,
that objects in our scale, i.e., ``macroscopic objects,'' can
partake all their properties is something that cannot be logically
proven. This assumption must be accepted as a metaphysical
hypothesis, and there are no known counterexamples to it.
Furthermore, this hypothesis was what Western philosophy has
preferred, from the Stoics to Leibniz's metaphysics.

Let us remember Leibniz's metaphysics' intuitive idea: \textit{no two individuals share all their properties; if they have the same attributes, they are not different, but the same individual.} This metaphysical principle was encapsulated in standard logic with the definition of identity given by Leibniz Law. This law says what we have expected: entities are identical if and only if they share all their properties, hence all their relations, that is, if and only if they are indistinguishable. 

What about the identity of properties? In standard logic, we usually say that two properties, $P$ and $Q$, are ``identical'' if they are satisfied by the same ``things.'' For instance, for Aristotle, the properties ``to be a human'' and ``to be a rational animal'' are ``identical'' in this sense. As an example from standard mathematics, consider the sets $\{ x\in \mathbb{R}|x^2-5x+6=0 \}$,  $\{ x\in \mathbb{N}| 1<x<4\}$, and $\{ x\in \mathbb{R}| x=2\vee x=3 \}$. These three sets are identical: they have the same extensions but different \emph{intensions.}\footnote{In technical terms, in extensional higher-order logics, we can define such a notion by saying that $P$ and $Q$ are identical when they have the same \textit{extensions}, that is, when they are satisfied by the same lower terms.} 

\textit{Classical} mathematical frameworks do not accommodate
indistinguishables; entities sharing all their attributes and being
just numerically distinct do not exist in classical mathematics (but
see below).  Individuals are unique, separable, at least in
principle, counted as one of a kind and presenting differences to
every \textit{other} object. There are no purely numerical identical
individuals: some form of Leibniz's Law holds. This is so within
standard logic and mathematics, and the ways of dealing with
indiscernibles require mathematical tricks such as confining them to
non-rigid structures\footnote{A structure (a domain comprising
relations over its elements) is rigid if its only automorphism
(bijections that preserve the relations of the structure) is the
identity function. Indiscernibility in a structure means that the
objects are invariant by some automorphism of the structure; in
rigid structures, an object is indiscernible just from itself.
Non-rigid (deformable) structures hide the object's identity so that
we may not be able to discern them by lack of distinctive relations
or properties. For details, see \cite[\S
6.5.2]{french_identity_2006}, \cite{kracoe05}.}. For example, take
the structure $\langle \mathbb{Z},+\rangle$, which represents the
integer numbers, $\mathbb{Z}$, and \emph{only} the standard addition
operation,  ``$+$.'' This structure is not rigid, since the
transformation $f(x)=-x$ is an automorphism of the structure, i.e.,
it keeps the individuals indiscernible within its point of view. To
see this, take the $2$ and $-2$. We cannot discern them
\emph{within} this structure. Imagine any property for $2$ defined
only with ``+,'' such as ``$2+1=3$.'' If we change the numbers by
the ``minus'' ones, we have ``$(-2)+(-1)= (-3)$.'' From within this
structure, the latter is identical to the former; we cannot
distinguish them. Of course, if we added additional properties to
the structure, such as the ``$<$'' relation, it would become rigid,
and we would be able to distinguish between $2$ and $-2$. However,
we cannot do it only with ``$+$.''

The search for \textit{legitimate} indiscernible
objects/individuals, in the above sense and without mathematical
tricks, requires a change of logic. We will retake this discussion
later on this paper, but we wish to turn to another kind of question
for now.

Some authors, such as Peter Geach, argue that identity is relative.
The only thing we can say, according to him, is that two individuals
$a$ and $b$ are (or not) identical relative to a sortal\footnote{A
sortal predicate enables to count the objects that obey the
predicate, such as ``being a philosopher.'' So, Isaac Newton and
Stephen Hacking would both be counted as ``Lucasian Professor of
Mathematics in Cambridge.''} predicate $F$; in the positive case, we
say that they are $F$-identical and can write $a =_F b$. In our
opinion, identity is absolute. Identity is, according to us, to be
associated with metaphysical identity, as explained above. It is
something an individual has that says that it is unique and, when it
appears in some other context, we are authorized to think that it is
the same individual that has appeared twice. Alternatively, an
individual's identity is its identity card, one for each individual:
it accompanies it in all contexts and, with its help, we can
distinguish the individual as being \textit{the same} individual of
a previous experience. Identity makes the individual's name a rigid
designator, denoting the same entity in all possible accessible
worlds.  As it is well known, David Hume guessed that there is no
such an identity; according to him, we recognize someone as being
\textit{the same} from a previous experience by habit, by
familiarity \cite[p.74 and \textit{passim}]{hume85}, but cannot
``logically'' prove that. Schr\"odinger had a similar opinion
regarding quantum entities when he says that
\begin{quote}
``[w]hen a familiar object reenters our ken, it is usually
recognized as a continuation of previous appearances, as being the
same thing. The relative permanence of individual pieces of matter
is the most momentous feature of both everyday life and scientific
experience. If a familiar article, say an earthenware jug,
disappears from your room, you are quite sure that somebody must
have taken it away. If after a time it reappears, you may doubt
whether it really is the same one $-$ breakable objects in such
circumstances are often not. You may not be able to decide the
issue, but you will have no doubt that the doubtful sameness has an
indisputable meaning $-$ that there is an unambiguous answer to your
query. So firm is our belief in the continuity of the unobserved
parts of the string!" \cite[p.204]{sch98}
\end{quote}
Entities partaking metaphysical identity are termed
\textit{individuals}. Can we think of  \textit{non-individuals} too?
If yes, can we give examples of entities of this kind? The first way
to think of them, by considering what we have said, is to deny them
the epithet ``to have an identity.'' What should it mean? The short
answer is that they would be entities sharing all their
characteristics, either substratum or properties and relations. From
now on, we shall avoid speaking of substratum and keep with bundle
theories \cite{teller_1998}. However, non-individuals, in our
formulation, are not simply metaphysically or numerically identical
entities, although  this is logically possible.\footnote{In his
criticism to the definition of identity given by Whitehead and
Russell in their \textit{Principia Mathematica} (Leibniz Law, in a
standard second-order language, $x = y := \forall F (Fx \lra Fy)$,
where $x$ and $y$ are individual terms and $F$ is a predicate
variable for individuals), F. P. Ramsey said precisely this: that we
could logically conceive entities violating the definition, sharing
all their properties, and even so not being the same entity
\cite[p.30]{ramsey65}.} Our notion is weaker, enabling
non-individuals to form collections (termed ``quasi-sets'') with
cardinalities greater than one so that no particular differences can
be ascribed to them. Furthermore, they would be indistinguishable
even if an omniscient demon (Laplace's demon) exchanged them with
one another; in this case, nothing would change in the world at all.
That is the difference: individuals, by definition, when permuted,
make a difference! This difference is of fundamental importance, for
it involves several other related notions which appear in physical
theories, such as space and time and, fundamentally, permutations.
We shall need to explain that further, but for now, we wish to
emphasize that we do not regard identity as something an entity
\textit{must} have. When something has an identity, then it is
absolute, it is metaphysical, and no two entities with identity can
be only numerically distinct. \textit{Non-individuals} are entities
that lack identity, that can be just numerically discerned, that
have all the same identity card. If one looks at one non-individual
here and there, one finds ``another'' one in a different context;
not even demons or gods will tell one if this new object is
``different'' or ``the same'' one found previously, as this would be
meaningless.

Nevertheless, once we think about more than one entity, one could
claim that they must be \textit{different}. Mathematically, this
would be expressed by the set-theoretical argument that once the
cardinal of a set is greater than one, its elements \textit{must} be
different. We stress that this depends on the set theory one is
taking into account. In standard set theories, such as the most
celebrated systems (the apparently most famous one is termed
``ZFC''),  this is true, but in \textit{quasi-set theory} (discussed
below), this is may not be the case. In quasi-set theory, we not
only can have collections (quasi-sets) of absolutely indiscernible
entities and with a cardinal greater than one, but we can also
quantify such ``non-individuals.'' Quasi-set theory shows that
Quine's motto of ``no entity without identity" \cite[p.23]{quine69}
does not hold in general, for even non-individuals can be values of
the variables of a regimented language.

\subsection{Identity in classical formal settings}
There is a problem concerning the metaphysical identity of the last
section: it cannot be defined in first-order languages
\cite{hodges83, french_identity_2006}.\footnote{First-order
languages deal with domains of individuals, their properties,
relations and operations over them. Quantified expressions like
``There exists some $x$ such that $\ldots$'' and ``For all
individuals $x$, $\ldots$'' applies only to individuals, and we
cannot say things like ``There is a relation among individuals
$\ldots$'' or ``For every property of individuals $\ldots$.'' In
logic, we say that first-order languages quantify over individuals
only.} We provide here a slightly technical explanation. As said
earlier, first-order languages speak of the individuals of some
domain. Usually, the axiomatizations take logical identity as
primitive (represented by a binary predicate ``=''), subject to
certain axioms (reflexivity and substitutivity). We can prove that
identity is an equivalence relation, really a congruence, whose
intended interpretation is the \ita{identity of the domain}; calling
it $D$, then we are referring to the set $\Delta_D := \{\langle
a,a\rangle : a \in D\}$, also called the \ita{diagonal of } $D$. But
it can be proven that there are other structures, called elementary
equivalent structures,\footnote{Elementary equivalent structures are
interpretations of a first-order language that preserve the same
truth sentences. From the language's point of view, one cannot
distinguish among such structures: they look the same.} which also
model ``='' but interprets this symbol in sets other than the
diagonal (op.cit.). So, within a first-order language, we never know
if we speak of the identity (or the difference) of two individuals
or of, say, classes of individuals.

Higher-order languages enable us to define logical identity by
Leibniz Law, but such logical identity is defined through
indiscernibility. If we wish to define indiscernibility instead, the
definition would be the same: agreement for all properties. So,
higher-order languages do not distinguish between these two
concepts. If we intend to speak of indiscernible but not identical
things, Leibniz Law does not help.\footnote{The distinction between
identity and indiscernibility can be made only in semantical terms;
see \cite{coskra97}.} Furthermore, if we aim to preserve some
meta-properties of our system (Henkin's completeness), we are
subject to find Henkin models so that two objects of the domain look
as indiscernible since they obey all the language's predicates, but
which are not the same element \cite[\S
6.3.2]{french_identity_2006}. In short, we need to conclude that
metaphysical identity cannot be defined. The most we can do is find
refuge in logical identity, but this, as we shall see soon, causes
troubles to quantum mechanics.

However, let us first put away the often-made claim that even
quantum objects can be discerned by spatio-temporal location.

\subsection{Identity and space and time}\label{subsect_idd-space-time}
There is still another way to look at identity in classical
settings: include space and time.  Orthodox non-relativistic quantum
mechanics makes use of \textit{classical} space and time or, as we
can say, ``Newtonian'' absolute notions. Intuitively, the classical
space and time structure is a space that looks, at least for small
regions, like the $\mathbb{R}^4$, namely three dimensions for space
($\mathbb{R}^3$) and one for time ($\mathbb{R}$). More precisely,
mathematically, the classical space-time is a manifold locally
isomorphic to $\mathbb{R}^4$, usually termed $\mathbb{E}^4$ (for
``Euclidean''); see \cite[Chap. 17]{penrose_road}.

This structure has some interesting features, but for us here, an
important characteristic is that it is a ``Hausdorff space.'' This
property of being Hausdorff means that, given any two points $a$ and
$b$, $a\neq b$, it is always possible to find two disjoint open sets
(say two open balls) $B_a$ and $B_b$ such that $a \in B_a$ and $b
\in B_b$. In extensional contexts, such as the ZFC set theory, a
property is confounded with a set; the objects that belong to the
set are precisely those satisfying the property. So, $a$ and $b$
have each a property not shared with the other, namely, to belong to
``its'' open set. Hence, Leibniz's Law applies, and they are
different. Notice that this holds for any \ita{two} objects $a$ and
$b$: once we have \ita{two}, they are distinct.  Therefore, we may
say that, within such a framework, there are no
indiscernibles!\footnote{In model theory, an important part of
logic, we can speak of ``indiscernibles'' in a sense, for instance,
\ita{Ramsey indiscernibles}. However, this is a way of speaking;
even these entities obey the classical theory of identity, therefore
being individuals. See \cite[Chap.15]{butwal18}.}

Let us see now how we can pretend to say that we have indiscernibles
within a classical framework.

\subsection{Indiscernibility in classical logical  settings}\label{sect_indisCS}
Still working in a classical setting, say the ZFC system, we can
mimic indiscernibility. In this subsection we expand the above
discussion about using non-rigid structures, presenting some of its
more technical concepts and ideas.

Usually, we say that the elements of a certain equivalence class are
indiscernible, and perhaps this is acceptable for certain purposes.
More technically, in doing that, we are restricted to a
\textit{non-rigid (or deformable) structure}. As we saw previously,
we say that a structure $\mathfrak{A} = \langle D, R_i \rangle, i
\in I$, is \textit{rigid} if its only automorphism is the identity
function; this means that we have a domain $D$, a non-empty set, and
a collection of relations over the elements of $D$, each one of a
certain arity $n=0,1,2,3,\ldots$.\footnote{That the identity mapping
is an automorphism is trivial. For all the argumentation, it is
enough to consider \textit{relational structures}, for distinguished
elements and operational symbols can be taken as particular kinds of
relations; also, we subsume all domains in just one.} If the
structure is not rigid, then it is is non-rigid or deformable. We
saw an example of a deformable structure earlier on, the $\langle
\mathbf{Z},+\rangle $. Another example of a deformable structure is
the field of the complex numbers, for the operation of taking the
conjugate is an automorphism. In such a structure $\mathfrak{C} =
\langle \mathbb{C}, 0, 1, +, \cdot \rangle$, the individuals $i$ and
$-i$ are indiscernible.

Given $\mathfrak{A}$ as above, we say that the elements $a$ and $b$
of $D$ are $\mathfrak{A}$-indiscernible if there exists $X \subseteq
D$ such that (i) for every automorphism $h$ of $\mathfrak{A}$, $h(X)
= X$, that is, $X$ is invariant by the automorphisms of the
structure, and (ii) $a \in X$ iff $b \notin X$. Otherwise, $a$ and
$b$ are $\mathfrak{A}$-discernible \cite{kracoe05}.

 It is clear that in a rigid structure, the only element indistinguishable from $a$ is $a$ itself since the only automorphism is the identity function. In informal parlance, we may say that $a$ and $b$ are $\mathfrak{A}$-indiscernible iff they are invariant by permutations that ``preserve the relations of the structure.''

Something like that is what we do in quantum mechanics. Roughly
speaking, the theory says that when we measure a certain observable
value for a quantum system in a certain state, the value does not
change before and after a permutation of particles of the same kind.
Physicists say that \ita{permutations are not observable}, and this
is expressed by the Indistinguishability Postulate.\footnote{In
technical terms, let us take a permutation $P$ between particles
denoted by $x_i$ and $x_j$. As usually stated, we may say that for
any $x_1, \ldots,x_n$,
\begin{equation}
    P(x_1,\ldots,x_i,\ldots,x_j,\ldots,x_n) \lra P(x_1,\ldots,x_j,\ldots,x_i,\ldots,x_n)
\end{equation}
 The \ita{Indistinguishabilility Postulate} is expressed in terms of ``expectation values;'' it says that
 \begin{equation}
     \langle \psi | \hat{A} | \psi \rangle = \langle P\psi | \hat{A} | P\psi \rangle
 \end{equation}
for any observable represented by a self-adjoint operator $\hat{A}$
and for any permutation operator $P$, being $\ket{\psi}$ the vector
state of the system.}

Leaving formal logic and mathematics for a while, let us consider
more general situations, which will lead us to a more detailed
discussion about quantum mechanics. We shall commence by emphasizing
the importance of the \ita{contexts}.

\section{Connecting identity to context}\label{seq:connecting-identity}
On many occasions, we are tempted to think about possible worlds
which are not actual. We wonder what our life would have been like
if we had taken different decisions at crucial moments. We can think
about an object, person, or animal, in many different circumstances,
which can differ from the actual ones. For example, suppose that we
have a pet cat and live in a small apartment. Given its living
conditions, the cat cannot catch the birds that he sees through the
window. He observes them with attention, craving for them but unable
to reach them. Thus, in our tiny-apartment world, our cat never
caught a bird. Furthermore, he never will because he cannot go out.
However, we can \textit{imagine} a different world, in which we live
in a house with a big yard in which our cat can wander out as many
times as it wants. In this big yard world, our cat can surely try to
catch a bird, and he will undoubtedly do so at least once.

The above story is an example of how we reason about
counterfactuals. We are tempted to conclude something that occurs in
a world that is not actual \emph{could} happen, even if that world
never becomes actual. This kind of reasoning is very natural in our
everyday life. However, what are the assumptions behind it? First,
somehow, our cat retains its identity among the different worlds:
the cat in the small apartment world is the same as the cat in the
big yard world. Both cats have the same name, color, same
capabilities, and desire to catch birds. Nevertheless, how can we
assure that the cat will retain its properties among the different
worlds? Perhaps, if we could afford a house with a big yard, we
could also afford fancy and tasty cat food. The cat gets used to it,
stays inside the house, and eats the whole day. In the fancy house
world, it might become idle to the point that it barely moves or
plays, as it happens with some cats. When it finally goes out to the
garden, it cannot catch birds anymore, as it became clumsy and slow.

The above example shows that we should not make hasty conclusions:
the properties of an object, person, or animal, might depend
strongly on the \textit{context} in which we are considering them.
In the small apartment, humble life, with cheap food, our cat is
playful and agile: it has a high probability of catching a bird but
no bird to catch. In the big house, those properties may or may not
be valid. The first lesson is: to assume that an object retains its
properties among different and incompatible worlds is not granted.
Even more so, one may ask: in which sense are the two cats in
different worlds the same? From a strict point of view, one may say
that the agile cat from our actual world is not the same as the idle
cat of the alternative reality. In the same way, we should not mix
the different worlds with counterfactual reasoning. If we conclude,
by studying our cat in this actual world, that he is very skilled in
chasing birds, we cannot use empirical information from our world to
conclude that the cat will indeed chase a bird in the alternative
world.

Thus, we are introduced to a profound philosophical problem by
thinking about the above straightforward situation: what are the
principles or conditions that grant identity to objects considered
in different possible worlds? Are we entitled to say that a given
object retains its identity when considered in different and
incompatible situations? Of course, in many situations of our daily
life, assuming that objects retain their identities and properties
in different contexts will work. Our bike works well on sunny and
rainy days and in diverse landscapes (such as cities or mountains).
Many characteristics of our bike -- such as its color or its range
of velocities -- are, to a great extent, \textit{context
independent}. However, we should not take this context independence
for granted. This is more so if we consider quantum systems that
define phenomena that lie far beyond our everyday experience. The
realm of the atom extends far beyond the {\aa}ngstr\"{o}m scale (ten
to the minus ten meters, which is something like $0,0000000001$
meters for one {\aa}ngstr\"{o}m!). The principles -- whatever they
are -- that allow us to identify properties and objects among
incompatible situations may no longer be valid for atomic systems.
Moreover, this seems to be the case, as the GHZ example above and
the following example show.

Suppose that Alice and Bob have separated labs, $L_{A}$ and $L_{B}$,
in which they perform their experiments. At a given time, a third
party prepares a quantum system capable of affecting what happens in
$L_{A}$ and $L_{B}$. Suppose that Alice decides to make an
experiment $P_{A}$ in her lab, in order to interact with the given
quantum system, and that Bob can do $P_{B}$ or $P'_{B}$ in $L_{B}$.
Due to the peculiarities of quantum mechanics, $P_{B}$ and $P'_{B}$
cannot be performed at the same time -- they are
\textit{incompatible} experiments. To understand what
\textit{incompatible} means, imagine the following situation: in
order to perform $P_{B}$, Bob must align a magnet in a given
direction $d$, and in order to perform $P'_{B}$, he must align its
magnet in a different direction $d'$. A magnet cannot point in two
different directions -- similarly, a clock's handle cannot point at
two different angles simultaneously. Thus, there are two
incompatible situations: either Alice performs experiment $P_{A}$
and Bob performs $P_{B}$, or Alice performs $P_{A}$ and Bob
$P'_{B}$. The two possibilities \textit{cannot} coexist in the same
world. Let us call these possibilities $W_{1}$ and $W_{2}$,
respectively.

Suppose now that Alice and Bob are in the process of deciding what
to do. They wonder about the experiments' possible outcomes in the
different situations, $W_{1}$ and $W_{2}$. Notice that they do not
need actually to perform the experiments. It is all about reasoning
in various alternatives without actually performing them. Now we
question: what is the status of the possible results of experiment
$P_{A}$ concerning $W_{1}$ and $W_{2}$? After the discussion about
the cat, we should not be as quick to identify what happens in
$W_{1}$ with $W_{2}$, even if we are talking about the same
experiment, $P_{A}$. In both possible worlds, Alice will perform the
same actions (she will orient the magnets in the same directions,
prepare the same reading apparatus, and so on). Is she going to
obtain the same results? What enables us to conclude that she will?
Notice that we are not asking here about an \textit{influence} of
Bob's actions in Alice ones: the laboratories can be very far away
in space and time. We are asking here whether we are entitled to
assume that there is some trace of identity among the results
obtained in different (and incompatible \textit{worlds}). As
expected, the answer is: no, we are not. Contradictions can be
readily achieved if we do so, as the cat and contextuality examples
suggest (and shown in technical research on quantum theory).

The actions required for experimenting $P_{A}$ are the same in
$W_{1}$ and $W_{2}$. Can we say that $P_{A}$ in $W_{1}$ is the same
as $P_{A}$ in $W_{2}$? After the cat discussion, let us be
conservative about the answer. We will say that $P_{A}$ in $W_{1}$
is \textit{indistinguishable} from $P_{A}$ in $W_{2}$. The two
experiments are completely alike: Alice will execute the same
actions in a system prepared with an equivalent procedure in both
worlds. However, we should not be tempted to claim they are the
same. The more so, we should not expect the same results. In this
sense, we say that the properties studied by experiment $P_{A}$ in
$W_{1}$ are indistinguishable from the properties studied by $P_{A}$
in $W_{2}$. We denote these properties by the pairs $(P_{A};W_{1})$
and $(P_{A};W_{2})$ and write $(P_{A};W_{1})\equiv (P_{A};W_{2})$,
to stress the fact that they are indistinguishable (but not
identical). A natural, logical formalism for describing this kind of
indistinguishability is the quasi-set theory. This theory allows us
to consider properties or objects in alternative worlds as
collections of indiscernible ur-elements.

If world $W_{1}$ becomes actual, Alice and Bob will perform their
actions, obtain their results, and record them. Out of these
results, what conclusions should they take about the possible
results associated with $W_{2}$? Are they entitled to reason in a
counterfactual way and combine the results of worlds $W_{1}$ and
$W_{2}$ to extract conclusions about them? Much caution should be
taken here, as the cat and contextual examples show. In principle,
there is no \textit{a priori} reason to do so. That we are allowed
to do so in many (but not all!) everyday situations is more a lucky
strike that we share with other creatures in our macroscopic reality
than a general rule. Counterfactual reasoning simplifies our
existence, but we should not expect it to be valid in every
situation. This lack of validity seems empirically suggested at
microscopic scales, which are very different from our own.

To summarize, we can state the following:
\begin{itemize}
    \item Even if state preparations and measurement procedures are completely alike among different worlds, we should not treat them as identical. In this sense, we speak about things such as indistinguishable properties and objects.
    \item Even if two experiments are completely indistinguishable, we should not expect the same results in different worlds.
    \item We should not derive conclusions from counterfactual reasoning, especially in the quantum domain. Such conclusions are not reliable and are not metaphysically justified.
\end{itemize}

\section{Quantum mechanics in classical logical settings}

In this section, we briefly review how the standard quantum
formalism performs the trick of treating indiscernible quantum
systems within the scope of classical logic (encompassing
mathematics). In doing so, we lay the groundwork for alternative
logics and mathematics, which provide an adequate description from
our perspective.

A glance at standard textbooks on quantum mechanics reveals that
they use classical mathematics, hence classical logic. However, the
claim that quantum mechanics requires a different logic, known as
quantum logic, can also often be found.\footnote{The field of
``quantum logic'' arose from Birkoff and von Neumann's 1936 seminal
paper. The reader interested in the subject is referred to the
following excellent papers: \cite{dalla04} and \cite{svozilQL}.}
These two observations seem contradictory. Why is this apparent
contradiction present in the literature?

The reason may be as follows. Most physicists are concerned with
physical problems being solved by quantum theory and not with
philosophical or logical foundational questions about it. Although
they might endorse some particular interpretation of quantum
mechanics, thus presupposing some concern with quantum theory's
philosophy, most physicists use ``classical'' mathematics in an
almost instrumentalist way. Thus, when dealing with entities that
would be indistinguishable, physicists use some mathematical tricks
to hide the identifications typical of our standard mathematical
languages. Let us see how they do it.

First, we recall that, in quantum mechanics' standard formulation, a
system's state is represented mathematically by a vector in a
Hilbert space. This vector, also called the wave function, is
supposed to encode all information available for that system in a
specific situation. Observables, which represent possible
experimental procedures and their outcomes, are self-adjoint
operators in the Hilbert space. When an observable is measured, the
state-vector enters (or ``collapse'') into one of the observable
operator's eigenvectors. Since this process is ``mysterious,'' in
the sense that the formalism does not explain how it happens, many
physicists try to avoid it, adopting alternative explanations.
Nevertheless, the primary mathematical object in quantum theory is
the Hilbert space and vectors in it. So, the question is how to
represent indistinguishable objects using the mathematics of
vectors.

Quantum particles come in two types: bosons and fermions. Their main
difference comes from their statistics: bosons follow the
Bose-Einstein statistics, whereas Fermions satisfy the Fermi-Dirac
one. Both statistics count objects as if they were
indistinguishable, contrary to the classical Maxwell-Boltzman
statistics.

Bosons are a typical type of indistinguishable quantum entities.
Bosons are a kind of quantum ``particles,'' and they are entirely
indistinguishable when prepared in the same quantum state. This
state is such that they share all the relevant quantum properties. A
system composed of, say, two bosons $1$ and $2$ in two possible
situations $A$ and $B$ is described by a symmetric wave function
such as the following.
\begin{equation}
    \Psi = \frac{1}{\sqrt{2}}\Big(\psi_1^A \psi_2^B + \psi_2^A \psi_1^B\Big),
\end{equation}
where $\psi_1^A \psi_2^B$ means system $1$ in the state $A$ and
system $2$ in $B$ and similarly for the other term. The
$\frac{1}{\sqrt{2}}$ is just a normalization factor required by the
formalism. $\Psi$ is invariant under the permutation of $1$ and $2$.
This invariance means that exchanging particle $1$ by $2$ (and
vice-versa) does not affect the state of the system. Consequently,
any measurement results are maintained under permutations.

This symmetrization of the wave function works, but it is a trick.
We are still using labels to ``name'' the particles because our
language and mental models have a hard time thinking otherwise. In
other words, this trick assumes, upfront, that bosons are
individuals. Suddenly, as if a miracle happened, permutations do not
conduce to different situations. However, this invariance was put
there by hand. We could give more detailed arguments as to why this
is a mathematical trick that does not make bosons indistinguishable,
but we hope the above example is sufficient for the reader to grasp
the main idea.

The use of the above trick is similar to confining the discussion to
a deformable (non-rigid) structure, as explained earlier. However,
as mentioned, within such classical settings, we can always go
``outside'' of the structure and identify the particles. This
possibility of identification is at odds with the hypothesis that
they are indiscernible.\footnote{The way to ``go outside'' the
quantum formalism is to go to the set-theoretical universe since all
mathematics used in quantum mechanics can be performed in terms of
sets.}

There is no way to escape this conclusion. As we have said before,
standard mathematics and logic are theories of individuals. This is
so for historical reasons: classical logic, mathematics, and even
classical physics were built with individuals in mind. Quantum
mechanics, of course, came to challenge those ideas and to question
the concepts of individuality.


\section{Alternative logical approaches\label{seq:logical-approaches}}

Assuming that indiscernibility is a core notion in quantum
mechanics, we should look for an alternative logical and
mathematical basis that considers it right from the start. This
bottom-up approach would not mimic it within a standard framework
from a top-bottom one. Our strategy is grounded in a metaphysics of
non-individuals (for detail, see \cite{french_identity_2006},
\cite{kraarebue20}, and references therein). Moreover, it tries to
develop mathematics compatible with such metaphysics. Consequently,
Schr\"{o}dinger logics and quasi-set theory were developed in the
1990s. Although they are mathematical developments independent of
the interpretations, the intended one is precisely to cope with such
non-individual entities. In this section, we will give a rough idea
about how quasi-set theory works. For a review about Schr\"{o}dinger
logics, see \cite[chap.8]{french_identity_2006}.


\subsection{Quasi-set theory}\label{qsettheory}

 In the quasi-theory $\mathfrak{Q}$, indiscernibility is a primitive concept, formalized by a binary relation ``$\equiv$'' satisfying the properties of an equivalence relation, but not full substitutivity.\footnote{If we add substitutivity to the postulates, then no differences between indiscernibility and logical first-order identity would be made.} In this notation, ``$x\equiv y $'' is thought to mean ``$x$ is indiscernible from $y$.'' This binary relation is a partial congruence in the following sense: for most relations, if $R(x,y)$ and $x \equiv x^\prime$, then $R(x^\prime, y)$ as well (the same holds for the second variable). The only relation to which this result does not hold is membership: $x \in y$ and $x^\prime \equiv x$ does not entail that $x^\prime \in y$; details in \cite{french_identity_2006,french_remarks_2010}).


 Quasi-sets can have as elements other quasi-sets, particular quasi-sets termed \ita{sets} which are copies of the sets in a standard theory (in the case, the Zermelo-Fraenkel set theory with the Axiom of Choice), and two kinds of atoms (entities which are not sets), termed $M$-atoms ($M$-objects), which are copies of a standard set theory with atoms (ZFA) and $m$-atoms ($m$-objects), which have the quanta as their intended interpretation, to whom it is supposed that the logical identity does not apply. If we eliminate the $m$-atoms, we are left with a copy of ZFA, the Zermelo-Fraenkel set theory with atoms. Hence, we can reconstruct all standard mathematics within $\mathfrak{Q}$ in such a ``classical part'' of the theory.

Functions cannot be defined in the standard way. When $m$-atoms are
present, it cannot distinguish between indiscernible arguments or
values. Therefore, the theory generalizes the concept to
``quasi-functions,'' which map indiscernible elements into
indiscernible elements. See below for more on this point.

Cardinals (termed ``quasi-cardinals,'' $qc$) are also taken as
primitive, although they can be proven to exist for finite qsets
(finite in the usual sense \cite{domenech_discussion_2007,
arenhart}).  The concept of quasi-cardinals can be used to speak of
``several objects.'' So, when we say that we have two indiscernible
q-functions, according to the above definition, we are saying that
we have a qset whose elements are indiscernible q-functions and
whose q-cardinal is two.\footnote{Quasi-cardinals turn to be
\textit{sets}, so we can use the equality symbol among them. We use
the notation $qc(x)=n$ (really, $qc(x) =_E n$, see below) for a
quasi-set $x$ whose cardinal is $n$.}. The same happens in other
situations.

An interesting fact is that qsets composed of several
indistinguishable $m$-atoms do not have an associated ordinal. This
lack of an ordinal means that these elements cannot be counted since
they cannot be ordered. However, we can still speak of a
collection's cardinal, termed its \textit{quasi-cardinal} or just
its \textit{q-cardinal}. This existence of a cardinal but not of an
ordinal is similar to what we have in QM when we say that we have
some quantity of systems of the same kind but cannot individuate or
count them, e.g., the six electrons in the level 2$p$ of a Sodium
atom.\footnote{To count a finite number of elements, say 4, is to
define a bijection from the set with these elements to the ordinal
$4 = \{0,1,2,3\}$. This counting requires that we identify the
elements of the first set.}

Identity (termed \textit{extensional identity}) ``$=_E$'' is defined
for qsets having the same elements (in the sense that if an element
belongs to one of them, then it belongs to the
another)\footnote{There are subtleties that require us to provide
further explanations. In $\mathfrak{Q}$, you cannot do the maths and
decide either a certain $m$-object belongs or not to a qset; this
requires identity, as you need to identify the object you are
referring to.

In quasi-set theory, however, one can hypothesize that \textit{if} a
specific object belongs to a qset, then so and so. This is similar
to Russell's use of the axioms of infinite ($I$)  and choice ($C$)
in his theory of types, which assume the existence of certain
classes that cannot be constructed, so going against Russell's
constructibility thesis. What was Russell's answer? He transformed
all sentences $\alpha$ whose proofs depend on these axioms into
conditionals of the form $I \to \alpha$ and $C \to \alpha$. Hence,
\textit{if} the axioms hold, \textit{then} we can get $\alpha$. We
are applying the same reasoning here: \textit{if} the objects of a
qset belong to the another and vice-versa, \textit{then} they are
extensionally identical. It should be noted that the definition of
extensional identity holds only for sets and $M$-objects.} or for
$M$-objects belonging to the same qsets. It can be proven that this
identity has all the properties of classical logical identity for
the objects to which it applies. However, it does not make sense for
$q$-objects. That is, $x =_E y$ does not have any meaning in the
theory if $x$ and $y$ are $m$-objects. It is similar to speak of
categories in the Zermelo-Fraenkel set theory (supposed consistent).
The theory cannot capture the concept, yet it can be expressed in
its language. From now on, we shall abbreviate ``$=_E$'' by ``$=$,''
as usual.

The postulates of $\mathfrak{Q}$ are similar to those of ZFU, but by
considering that now we may have $m$-objects. The notion of
indistinguishability is extended to qsets through an axiom that says
that two qsets with the same q-cardinal and having the same
``quantity'' (we use q-cardinals to express this) of elements of the
same kind (indistinguishable among them) are indiscernible too. As
an example, consider the following: two sulfuric acid molecules
H$_2$SO$_4$ are seen as indistinguishable qsets, for both contain
q-cardinal equals to 7 (counting the atoms as basic elements), and
the elements of the sub-collections of elements of the same kind are
also of the same q-cardinal (2, 1, and 4 respectively). Then we can
state that ``H$_2$SO$_4$ $\equiv$ H$_2$SO$_4$,'' but of course, we
cannot say that ``H$_2$SO$_4$ $=$ H$_2$SO$_4$,'' as for in the
latter, the two molecules would not be two at all, but just the same
molecule (supposing, of course, that ``$=$'' stands for classical
logical identity). In the first case, notwithstanding, they count as
two, yet we cannot say which is which.

Let us speak a little bit more about quasi-functions. Since
physicists and mathematicians may want to talk about random
variables over qsets as a way to model physical processes, it is
important to define functions between qsets. This can be done
straightforwardly, \label{qfunctions} and here we consider binary
relations and unary functions only. Such definitions can easily be
extended to more complicated multi-valued functions. A (binary)
q-relation between the qsets $A$ and $B$ is a qset of pairs of
elements (sub-collections with q-cardinal equals 2), one in $A$, the
other in $B$.\footnote{We are avoiding the long and boring
definitions, as, for instance, the definition of ordered pairs,
which presuppose lots of preliminary concepts, just to focus on the
basic ideas. For details, the interested reader can see the
indicated references.} Quasi-functions (q-functions) from $A$ to $B$
are binary relations between $A$ and $B$ such that if the pairs
(qsets) with $a$ and $b$ and with $a^\prime$ and $b^\prime$ belong
to it and if $a \equiv a^\prime$, then $b \equiv b^\prime$ (with
$a$'s belonging to $A$ and the $b$'s to $B$). In other words, a
q-function maps indistinguishable elements into indistinguishable
elements. When there are no $m$-objects involved, the
indistinguishability relation collapses in the extensional identity,
and the definition turns to be equivalent to the classical one. In
particular, a q-function from a ``classical'' set such as $\{1,-1\}$
to a qset of indiscernible q-objects with q-cardinal $2$ can be
defined so that we cannot know which q-object is associated with
each number (this example will be used below).

To summarize, in this section, we showed that the concept of
indistinguishability, which conflicts with Leibnitz's Principle of
the Identity of Indiscernibles, can be incorporated as a
metaphysical principle in a modified set theory with
indistinguishable elements. This theory contains ``copies'' of the
Zermelo-Frankel axioms with \textit{Urelemente} as a particular case
when no indistinguishable $q$-objects are involved. This theory will
provide us the mathematical basis for formally talking about
indistinguishable properties, which we will show can be used in a
theory of quantum properties.  We will see in the next section how
we can use those indistinguishable properties to avoid
contradictions in quantum contextual settings such as KS.


\section{Formulating quantum mechanics within quasi-set theory\label{seq:formulating}}

As we have seen, the quasi-set theory enables us to form collections
(the quasi-sets) of ``absolutely'' indiscernible elements. In this
theory, even if one goes outside the relevant structures, they will
not become rigid: this mathematical universe is not rigid. Thus, the
quasi-set theory is a suitable device to develop a quantum theory
where indiscernibility is considered from the start as a fundamental
notion. This section explains how quantum mechanics (in the Fock
space formalism) can be developed within the quasi-set theory \qst.
The current development is based in \cite{domenech_q-spaces_2008}
and is technical. This level of mathematical formality is necessary
to provide essential details. The reader unconcerned with such
technicalities may skip this section and proceed directly to the
conclusions.

\subsection{The \qst-spaces}\label{qsp}

In the standard mathematical formalisms, the assumptions that
quantum entities of the same kind must be indiscernible are hidden
behind mathematical tricks such as symmetrizing wave-functions and
vectors. In order to avoid these tricks, we introduce the notion of
\qst-spaces. The resulting framework is termed  \textit{nonreflexive
quantum mechanics or, simply, nonreflexive}.

We begin with a q-set of real numbers $\epsilon =
\{\epsilon_{i}\}_{i \in I }$, where $I$ is an arbitrary collection
of indexes, denumerable or not. Since it is a collection of real
numbers, which may be constructed in the classical part of \qst, we
have that $Z(\epsilon)$. Intuitively, the elements $\epsilon_{i}$
represent the eigenvalues of a physical observable $\hat{O}$, that
is, they are the values such that
$\hat{O}|\varphi_{i}\rangle=\epsilon_{i}|\varphi_{i}\rangle$, with
$|\varphi_{i}\rangle$ the corresponding eigenstates. Since
observables are Hermitian operators, the eigenvalues are real
numbers. Thus, we are justified in assuming that elements of
$\epsilon$ are real numbers. Consider then the quasi-functions
$f:\epsilon \longrightarrow \mathcal{F}_{p}$, where
$\mathcal{F}_{p}$ is the quasi-set formed of all finite and pure
quasi-sets (that is, finite quasi-sets whose only elements are
indistinguishable m-atoms). Each of these $f$ is a q-set of ordered
pairs $\langle \epsilon_{i}, x\rangle$ with
$\epsilon_{i}\in\epsilon$ and $x\in\mathcal{F}_{p}$. From
$\mathcal{F}_{p}$ we select those quasi-functions $f$ which
attribute a non-empty q-set only to a finite number of elements of
$\epsilon$, the image of $f$ being $\emptyset$ for the other cases.
We call $\mathcal{F}$ the quasi-set containing only these
quasi-functions. Then, the quasi-cardinal of most of the q-sets
attributed to elements of $\epsilon$ according to these
quasi-functions is $0$. Now, elements of $\mcal{F}$ are
quasi-functions which we read as attributing to each $\epsilon_{i}$
a q-set whose quasi-cardinal we take to be the occupation number of
this eigenvalue. We write these quasi-functions as
$f_{\epsilon_{i_{1}}\epsilon_{i_{2}}\ldots\epsilon_{i_{m}}}$.
According to the given  intuitive interpretation, the levels
$\epsilon_{i_{1}}\epsilon_{i_{2}}\ldots\epsilon_{i_{m}}$ are
occupied. We say that if the symbol $\epsilon_{i_{k}}$ appears
$j$-times, then the level $\epsilon_{i_{k}}$ has occupation number
$j$. For example, the notation
$f_{\epsilon_{1}\epsilon_{1}\epsilon_{1}\epsilon_{2}\epsilon_{3}}$
means that the level $\epsilon_{1}$ has occupation number $3$ while
the levels $\epsilon_{2}$ and $\epsilon_{3}$ have occupation numbers
$1$. The levels that do not appear have occupation number zero.
Another point to be remarked is that since the elements of
$\epsilon$ are real numbers, we can take the standard ordering
relation over the reals and order the indexes according to this
ordering in the representation
$f_{\epsilon_{i_{1}}\epsilon_{i_{2}}\ldots\epsilon_{i_{m}}}$.
\label{or} This will be important when we consider the cases for
bosons and fermions.

The quasi-functions of $\mcal{F}$ provide the key to the solution to
the problem of labeling states. Since we use pure quasi-sets as the
images of the quasi-functions, there is simply no question of
indexes for particles, for all that matters are the quasi-cardinals
representing the occupation numbers. To make it clear that
permutations change nothing, one needs only to notice that a
quasi-function is a q-set of weakly ordered pairs.\footnote{A weak
ordered pair is a qset having just one element (that is, its
cardinal is one). We cannot name such an element, for we need an
identity to do that. SO, it can be taken as \textit{one} element of
a kind.} Taking two of the pairs belonging to some quasi-function,
let us say $\langle\epsilon_{i}, x \rangle$, $\langle \epsilon_{j},
y\rangle$, with both $x$ and $y$ non-empty, a permutation of
particles would consist in changing elements from $x$ with elements
from $y$. However, by the unobservability of permutations
theorem,\footnote{This theorem says that if we exchange an element
of a qset by an indistinguishable one, the resulting qset turns to
be indistinguishable from the original one.} what we obtain after
the permutation is a q-set indistinguishable from the one we began
with. Remember also that a quasi-function attributes
indistinguishable images to indistinguishable items; thus, the
indistinguishable q-set resulting from the permutations will also be
in the image of the same eigenvalue. To show this point precisely,
we recall that by definition $\langle\epsilon_{i}, x\rangle$
abbreviates $[[\epsilon_{i}],[\epsilon_{i},x]]$,\footnote{We are
leaving aside the subindices in this notation.} and an analogous
expression holds for $\langle\epsilon_{j}, y\rangle$. Also, by
definition, $[\epsilon_{i},x]$ is the collection of all the items
indistinguishable from $\epsilon_{i}$ or from $x$ (taken from a
previously given q-set). For this reason, if we permute $x$ with
$x'$, with $x\equiv x'$ we change nothing for $[\epsilon_{i},x]
\equiv [\epsilon_{i},x']$. Thus, we obtain
$\langle\epsilon_{i},x\rangle \equiv \langle\epsilon_{i},x'\rangle$
and the ordered pairs of the permuted quasi-function will be
indiscernible (the same if there are no m-atoms involved). Thus, the
permutation of indistinguishable elements does not produce changes
in the quasi-functions.

\subsection{A Vector Space Structure}

Now, we wish to have a vector space structure to represent quantum
states. To do that, we need to define addition and multiplication by
scalars. Before we go on, we must notice that we cannot define these
operations directly on the q-set $\mcal{F}$, for there is no simple
way to endow it with the required structure; our strategy here is to
define $\star$ (multiplication by scalars) and $+$ (addition of
vectors) in a q-set whose vectors will be quasi-functions from
$\mcal{F}$ to the set of complex numbers $\mathbb{C}$. Let us call
$C$ the collection of quasi-functions that assign to every $f\in
\mathcal{F}$ a complex number. Once again, we select from $C$ the
sub-collection $C_{F}$ of quasi-functions $c$ such that every $c\in
C_{F}$ attributes complex numbers $\lambda \neq 0$ for only a finite
number of $f\in \mathcal{F}$. Over $C_{F}$, we can define a sum and
a product by scalars in the same way as it is usually done with
functions as follows.
\begin{dfn}
Let $\gamma$ $\in \mcal{C}$, and $c$, $c_{1}$ and $c_{2}$ be
quasi-functions of $C_{F}$, then
$$(\gamma\star c)(f) \igual \gamma(c(f))$$
$$(c_{1}+ c_{2})(f) \igual  c_{1}(f) + c_{2}(f)$$
\end{dfn}
The quasi-function $c_{0}\in C_{F}$ such that $c_{0}(f)= 0$ for
every $f\in \mcal{F}$ acts as the null element for the sum
operation. This can be shown as follows:
\begin{equation}
(c_{0}+c)(f)= c_{0}(f)+c(f)= 0+c(f)= c(f), \forall f.
\end{equation}
With both the operations of sum and multiplication by scalars
defined as above we have that $\langle C_{F},\mathbb{C}, +,\star
\rangle$ has the structure of a complex vector space, as one can
easily check. Some of the elements of $C_F$ have a special status
though; if $c_{j}\in C_{F}$ are the quasi-functions such that
$c_{j}(f_{i})= \delta_{ij}$ (where $\delta_{ij}$ is the Kronecker
symbol), then the vectors $c_{j}$ are called the basis vectors,
while the others are linear combinations  of them. For notational
convenience, we can introduce a new notation for the q-functions in
$C_{F}$; suppose $c$ attributes a $\lambda \not= 0$ to some $f$, and
$0$ to every other quasi-function in $\mcal{F}$. Then, we propose to
denote $c$ by $\lambda f$. The basis quasi-functions will be denoted
simply $f_{i}$, as one can check. Now, multiplication by scalar
$\alpha$ of one of these quasi-functions, say $\lambda f_{i}$ can be
read simply as $(\alpha \cdot \lambda) f_{i}$, and sum of
quasi-functions $\lambda f_{i}$ and $\alpha f_{i}$ can be read as
$(\alpha + \lambda) f_{i}$. What about the other quasi-functions in
$C_{F}$? We can extend this idea to them too, but with some care:
if, for example $c_{0}$ is a quasi-function such that $c_{0}(f_{i})=
\alpha$ and $c_{0}(f_{j})= \lambda$, attributing $0$ to every other
quasi-function in $\mcal{F}$, then $c_{0}$ can be seen as a linear
combination of quasi-functions of a basis; in fact, consider the
basis quasi-functions $f_{i}$ and $f_{j}$, (this is an abuse of
notation, for they are representing quasi-functions in $C_F$ that
attribute 1 to each of these quasi-functions). The first step
consists in multiplying them by $\alpha$ and $\lambda$,
respectively, obtaining $\alpha f_{i}$ and $\lambda f_{j}$ (once
again, this is an abuse, for these are quasi-functions in $C_F$ that
attribute the mentioned complex numbers to $f_{i}$ and to $f_{j}$).
Now, $c_{0}$ is in fact the sum of these quasi-functions, that is,
$c_{0} = \alpha f_{i} + \lambda f_{j}$, for this is the function
which does exactly what $c_{0}$ does. One can then extend this to
all the other quasi-functions in $C_F$ as well.

\subsection{Inner Products}

The next step in our construction is to endow our vector space with
an inner product. This is a necessary step for we wish to calculate
probabilities and mean values. Following the idea proposed in
\cite{domenech_q-spaces_2008}, we introduce two kinds of inner
products, which lead us to two Hilbert spaces, one for bosons and
another for fermions. We begin with the case for bosons.
\begin{dfn}
Let $\delta_{ij}$ be the Kronecker symbol and
$f_{\epsilon_{i_{1}}\epsilon_{i_{2}}\ldots\epsilon_{i_{n}}}$ and
$f_{\epsilon_{i'_{1}}\epsilon_{i'_{2}}\ldots\epsilon_{i'_{m}}}$ two
basis vectors (as discussed above), then
\begin{equation}
f_{\epsilon_{i_{1}}\epsilon_{i_{2}}\ldots\epsilon_{i_{n}}}\circ
f_{\epsilon_{i'_{1}}\epsilon_{i'_{2}}\ldots\epsilon_{i'_{m}}} \igual
\delta_{nm}\sum_{p}\delta_{i_{1}pi'_{1}}\delta_{i_{2}pi'_{2}}\ldots\delta_{i_{n}pi'_{n}}.
\end{equation}
\end{dfn}
Notice that this sum is extended over all the permutations of the
index set $i'=(i'_{1},i'_{2},\ldots,i'_{n})$; for each permutation
$p$, $pi'=(pi'_{1},pi'_{2},\ldots,pi'_{n})$.

For the other vectors, the ones that can be seen as linear
combinations in the sense discussed above, we have
\begin{equation}
(\sum_{k}\alpha_{k}f_{k})\circ(\sum_{k}\alpha'_{k}f'_{k}) \igual
\sum_{kj}\alpha_{k}^{\ast}\alpha'_{j}(f_{k}\circ f'_{j}),
\end{equation}
where $\alpha^{\ast}$ is the complex conjugate of $\alpha$. Now, let
us consider fermions. As remarked above in page \pageref{or}, the
order of the indexes in each
$f_{\epsilon_{i_{1}}\epsilon_{i_{2}}\ldots\epsilon_{i_{n}}}$ is
determined by the canonical ordering in the real numbers. Thus, we
define another $\bullet$ inner product as follows, which will do the
job for fermions.
\begin{dfn}
Let $\delta_{ij}$ be the Kronecker symbol and
$f_{\epsilon_{i_{1}}\epsilon_{i_{2}}\ldots\epsilon_{i_{n}}}$ and
$f_{\epsilon_{i'_{1}}\epsilon_{i'_{2}}\ldots\epsilon_{i'_{m}}}$ two
basis vectors, then
\begin{equation}
f_{\epsilon_{i_{1}}\epsilon_{i_{2}}\ldots\epsilon_{i_{n}}}\bullet
f_{\epsilon_{i'_{1}}\epsilon_{i'_{2}}\ldots\epsilon_{i'_{m}}} \igual
\delta_{nm}\sum_{p}\sigma_{p}\delta_{i_{1}pi'_{1}}\delta_{i_{2}pi'_{2}}\ldots\delta_{i_{n}pi'_{n}}
\end{equation}
where: $\sigma_{p} = 1$ if p is even and $\sigma_{p} = -1$ if p is
odd.
\end{dfn}
This definition can be extended to linear combinations as in the
previous case.

\subsection{Fock spaces using \qst-spaces}

We begin with a definition to simplify the notation. For every
function
$f_{\epsilon_{i_{1}}\epsilon_{i_{2}}\ldots\epsilon_{i_{n}}}$ in
$\mcal{F}$, we put
$$\alpha |\epsilon_{i_{1}}\epsilon_{i_{2}}\ldots\epsilon_{i_{n}}) \igual \alpha f_{\epsilon_{i_{1}}\epsilon_{i_{2}}\ldots\epsilon_{i_{n}}}$$
Note that this is a slightly modified version of the standard
notation. We begin with the case of bosons.

Suppose a normalized vector $|\alpha \beta \gamma \ldots )$, where
the norm is taken from the corresponding inner product. Let $\zeta$
stand for an arbitrary collection of indexes. We define
$a_{\alpha}^{\dagger}|\zeta ) \propto |\alpha \zeta )$ in such a way
that the proportionality constant satisfies $a_{\alpha}^{\dagger}
a_{\alpha} |\zeta ) = n_{\alpha} |\zeta )$. From this it will
follow, as usual, that:
$$((\zeta|a_{\alpha}^{\dagger})(a_{\alpha}|\zeta)) = n_{\alpha}.$$
\begin{dfn}
$a_{\alpha} | \ldots n_{\alpha} \ldots ) := \sqrt{n_{\alpha}} |
\ldots n_{\alpha} - 1 \ldots )$
\end{dfn}
On the other hand, $$a_{\alpha}a_{\alpha}^{\dagger} | \ldots
n_{\alpha} \ldots ) = K \sqrt{n_{\alpha} + 1}|\ldots n_{\alpha}
\ldots ),$$ where $K$ is a proportionality constant. Applying
$a_{\alpha}^{\dagger}$ again, we have
$$a_{\alpha}^{\dagger}a_{\alpha}a_{\alpha}^{\dagger} | \ldots
n_{\alpha} \ldots ) = K^2 \sqrt{n_{\alpha} + 1}|\ldots n_{\alpha} +
1 \ldots ).$$ Using the fact that $a_{\alpha}^{\dagger} a_{\alpha}
|\zeta ) = n_{\alpha} |\zeta )$, we have that
$$(a_{\alpha}^{\dagger}a_{\alpha})a_{\alpha}^{\dagger} | \ldots
n_{\alpha} \ldots ) = \sqrt{n_{\alpha} + 1}K|\ldots n_{\alpha} + 1
\ldots ).$$ So, $K = \sqrt{n_{\alpha} + 1}$. Then, we have
\begin{dfn}
$a_{\alpha}^{\dagger} | \ldots n_{\alpha} \ldots ) :=
\sqrt{n_{\alpha} + 1}| \ldots n_{\alpha} + 1 \ldots ).$
\end{dfn}
From this definition, with additional computations, we obtain
$(a_{\alpha}a_{\beta}^{\dagger} -
a_{\beta}^{\dagger}a_{\alpha})|\psi) = \delta_{\alpha \beta}|\psi)$.
In our language, this means the same as $$[a_{\alpha} ;
a_{\beta}^{\dagger}] = \delta_{\alpha \beta} I.$$ In an analogous
way, it can be shown that
$$[a_{\alpha}; a_{\beta}] = [a_{\alpha}^{\dagger};a_{\beta}^{\dagger}] = 0.$$
So, the bosonic commutation relation is the same as in standard Fock
space formalism.

For fermionic states, we use the antisymmetric product
``$\bullet$.'' We begin by defining the creation operator
$C_{\alpha}^{\dagger}$.
\begin{dfn}
If $\zeta$ is a collection of indexes of non-null occupation
numbers, then $C_{\alpha}^{\dagger} := \alpha |\zeta)$
\end{dfn}
If $\alpha$ is in $\zeta$, then $|\alpha \zeta)$ is a vector of null
norm. This implies that $(\psi|\alpha \zeta) = 0$, for every $\psi$.
It follows that systems in states of null norm have no probability
of being observed. Furthermore, their addition to another vector
does not contribute to any observable difference. To take the
situation into account, we have the following definition.
\begin{dfn}
Two vectors $|\phi)$ and $|\psi)$ are similar if the difference
between them is a linear combination of null norm vectors. We denote
similarity of $|\phi)$ and $|\psi)$ by $|\phi) \cong |\psi)$.
\end{dfn}
Using the definition of $C_{\alpha}^{\dagger}$ we can describe what
is the effect of $C_{\alpha}$ over vectors, namely
$$(\zeta| C_{\alpha} := (\alpha \zeta|.$$

Then, for any vector $|\psi)$, $$(\zeta| C_{\alpha}|\psi) = (\alpha
\zeta|\psi) = 0$$ for $\alpha \in \zeta$ or $(\psi|\alpha \zeta) =
0$. Then, if $|\psi) = |0)$, then $(\zeta| C_{\alpha}|0) = (\alpha
\zeta|0) = 0$. So, $C_{\alpha}|0)$ is orthogonal to any vector that
contains $\alpha$, and also to any vector that does not contain
$\alpha$, so that it is a linear combination of null norm vectors.
So, we can put by definition that $\vec{0} := C_{\alpha}|0)$. In an
analogous way, if $ \sim \alpha$ denotes that $\alpha$ has
occupation number zero, then we can also write $C_{\alpha}|(\sim
\alpha) \ldots) = \vec{0}$, where the dots mean that other levels
have arbitrary occupation numbers.

Now, using our notion of similar vectors, we can write
$C_{\alpha}|0) \cong \vec{0}$ and $C_{\alpha}|(\sim \alpha) \ldots)
\cong \vec{0}$. The same results are obtained when we use $\cong$
and the sign of identity. By making $|\psi) = |\alpha)$, we have
$(\zeta| C_{\alpha}|\alpha) = (\alpha \zeta|\alpha) = 0$ in every
case, except when $|\zeta) = |0)$. In that case, $(0|
C_{\alpha}|\alpha) = 1$. Then, it follows that $C_{\alpha}|\alpha)
\cong 0$. In an analogous way, we obtain $C_{\alpha}|\alpha \zeta) =
\cong |(\sim \alpha) \zeta)$ when $\alpha \notin \zeta$. In the case
$\alpha \in \zeta$, $|\alpha \zeta)$ has null norm, and so, for
every $|\psi)$: $$(\alpha \zeta| C_{\alpha}^{\dagger}|\psi) =
(\alpha \zeta|\alpha \psi) = 0.$$ It then follows that
$$(\psi|C_{\alpha}|\alpha \zeta) = 0,$$
so that $C_{\alpha}|\alpha \zeta)$ has null norm too.

Now we calculate the anti-commutation relation obeyed by the
fermionic creation and annihilation operators. We begin calculating
the commutation relation between $C_{\alpha}$ and
$C_{\beta}^{\dagger}$. We do that by studying the relationship
between $|\alpha \beta)$ and $|\beta \alpha)$. Let us consider the
sum $|\alpha \beta) + |\beta \alpha)$. The product of this sum with
any vector distinct from $|\alpha \beta)$ is null. For the product
with $|\alpha \beta)$ we obtain $(\alpha \beta|[|\alpha \beta) +
|\beta \alpha)] = (\alpha \beta || \alpha \beta) + (\alpha \beta ||
\beta \alpha)$. By definition, this is equal to $\delta_{\alpha
\alpha} \delta_{\beta \beta} - \delta_{\alpha
\beta}\delta_{\beta\alpha} + \delta_{\alpha
\beta}\delta_{\alpha\alpha} - \delta_{\alpha
\alpha}\delta_{\beta\beta}$. This is equal to $1 - 0 + 0 - 1 = 0$.

The same conclusion holds if we multiply the sum $|\alpha \beta) +
|\beta \alpha)$ by $(\beta \alpha|$. It then follows that $|\alpha
\beta) + |\beta \alpha)$ is a linear combination of null norm
vectors, which we denote by $|nn)$, so that
$$|\alpha \beta) = - |\beta \alpha) + |nn).$$
Given that, we can calculate
$$C_{\alpha}^{\dagger}C_{\beta}^{\dagger}|\psi) = |\alpha \beta\psi)
= -|\beta \alpha |\psi) + |nn) = -
C_{\beta}^{\dagger}C_{\alpha}^{\dagger}|\psi) + |nn).$$ From this it
follows that $\{C_{\alpha}^{\dagger};C_{\beta}^{\dagger}\}|\psi) =
|nn)$. We do not lose generality by setting
$\{C_{\alpha}^{\dagger};C_{\beta}^{\dagger}\}|\psi) = 0 $. In an
analogous way we conclude that $$\{C_{\alpha};C_{\beta}\}|\psi) =
0.$$

Now we calculate the commutation relation between $C_{\alpha}$ and
$C_{\beta}^{\dagger}$. There are some cases to be considered. We
first assume that $\alpha \neq \beta$. If $\alpha \notin \psi$ or
$\beta \in \psi$ then $$\{C_{\alpha};C_{\beta}^{\dagger}\}|\psi)
\approx \vec{0}.$$ If $\alpha \in \psi$ and $\beta \notin \psi$,
assuming that $\alpha$ is the first symbol in the list of $\psi$,
then $\{C_{\alpha};C_{\beta}^{\dagger}\}|\psi) = C_{\alpha}|\beta
\psi) + C_{\beta}^{\dagger}|\psi(\sim \alpha))\cong -|\beta \psi
(\sim \alpha)) + |\beta \psi (\sim \alpha)) = \vec{0}$. If $\alpha =
\beta$ and $\alpha \in \psi$, then
$\{C_{\alpha};C_{\alpha}^{\dagger}\}|\psi) = C_{\alpha}|\alpha \psi)
+ C_{\alpha}^{\dagger}|\psi(\sim \alpha))\cong \vec{0} + |\psi) =
|\psi)$. If $\alpha = \beta$ and $\alpha \notin \psi$, then
$\{C_{\alpha};C_{\alpha}^{\dagger}\}|\psi) = C_{\alpha}|\alpha \psi)
+ C_{\alpha}^{\dagger}|\psi(\sim \alpha))\cong |\psi) + \vec{0} =
|\psi)$. In any case, we recover
$\{C_{\alpha};C_{\alpha}^{\dagger}\}|\psi) \cong \delta_{\alpha
\beta} |\psi)$. So, we can put
$$\{C_{\alpha};C_{\alpha}^{\dagger}\} = \delta_{\alpha \beta}.$$
It then follows that the commutation properties in \qst-spaces are
the same as in traditional Fock spaces.

Using this formalism, we can adapt all the developments done in
\cite[Chap.7]{mat67} and \cite[Chap.20]{mer70} for the number
occupation formalism. However, contrary to what happens in these
books, no previous (even unconscious) assumptions about quantum
objects' individuality is taken into account.

\section{Conclusions}
It is an exciting question to ask if we need to change logic every
time we find difficulties with the classical one. Are there other
ways to circumvent the problems, such as in the quantum case, using
the tricks mentioned above, or choosing an alternative
interpretation? This question makes sense. However, we think that
every theory, even a mathematical one, starts from metaphysical
hypotheses, even if not made explicit. We have stated above that
classical logic, standard mathematics, and classical physics were
developed with the classical enclosing world in our minds. This
world is one of individuals that have an identity. So, two of those
individuals cannot possibly be different.

Nevertheless, quantum mechanics brought us a different world, a
world with no proper names. In the quantum world, objects are (in
most cases) precisely alike, and permutations between objects of the
same kind do not lead to any physical differences. Here we emphasize
that it is not that these are not \ita{measurable} differences;
\ita{they are no differences at all.} So, we arrive at the following
conclusions.

\begin{description}
\item[1.] Indistinguishability is essential in quantum mechanics, regardless of interpretation. In our opinion, it should be placed at an equal level of importance in quantum foundations to concepts such as entanglement, contextuality, and nonlocality.

\item[2.] Ontological and epistemic aspects matter. Any physical theory is grounded in interpretations due to the possibility of associating different world views (or metaphysics) to a theory. Parodying Poincar\'e, we can say that physics is (also) a domain where we give the same name to distinct things.\footnote{Poincar\'e was referring to mathematics: ``mathematics is the art of giving the same name to distinct things" --- look at \cite{ver12}. Of course, he spoke within the framework of axiomatized mathematical theories, able to have different models.}

\item[3.] Since mathematics and logic need to reflect the assumed metaphysical aspects (we could speak in terms of ontology), quantum mechanics' formalism and physical theories should do the same.

\end{description}

Let us expand on this last point with an example involving logic.
It is common to say that in order to obtain intuitionistic logic, it
is enough to drop the excluded middle law from the axioms of
classical logic. From a purely formal point of view, this is
correct. However, logic is not only syntax. It also involves
semantic aspects and even pragmatic ones (making references to who
uses the logic and why). Let us consider semantics. Although
classical and intuitionistic logic differs syntactically just by one
axiom, semantically, they are much different. Classical
propositional logic can be described through truth-tables;
intuitionistic logic cannot. In classical logic, any proposition is
either true or false, yet we may not know what the case is; in
intuitionistic logic, the notions of true and false are different.
In this logic, a proposition $p$ is true if there is a ``process''
to get it, and false if a process for obtaining $p$ leads to a
contradiction. Other differences can be pointed out. For instance,
in classical logic, something exists if its nonexistence creates a
contradiction. In intuitionistic logic, something exists if it can
be created by our imagination.

This example shows that in order to consider a logic, semantical
aspects must at least be considered. Of course, this is true also
with physical theories. Otherwise, we risk having a purely
mathematical theory. However, what corresponds to semantics in the
quantum case? We chose interpretations because quantum mechanics, as
Yuri Manin wrote, ``does not really have its own language'' \cite[p.
84]{manin77}. At least not yet. Indeed, the standard formalism
grounded on Hilbert spaces makes use of the language of standard
functional analysis, which presupposes classical mathematics and
logic, with all the problems seem before (in regarding quantum
phenomena). A proper language should reflect the indiscernibility of
quanta from the start, without tricks!

As we showed in this paper, such a correct language can be
constructed. In this paper, we examined content and context in
quantum physics. We provided examples of context for the classical
and quantum realms and argued that the quantum situation is
fundamentally different. Furthermore, we reasoned that
context-dependency in the quantum world is intrinsically connected
to the lack of identity. Thus, the non-identity of individuals is an
essential feature of the quantum world. Since the standard
mathematics used in physics does not exactly allow for objects who
lack identity, i.e., indistinguishable objects, we advocated for
using a different mathematical structure in physics: quasi-set
theory. Quasi-set theory includes standard mathematic in it but also
contains indistinguishable objects. We believe that recreating
quantum physics in terms of quasi-set theory and its underlying
logic would result in thinking closer to a more reasonable ontology
for the quantum world than currently available ontologies. This way
of thinking may lead to exciting insights into quantum ontologies
and fundamental physical principles that define quantum mechanics.


\end{document}